\documentclass[sigconf]{acmart}

\usepackage{booktabs} 
\usepackage{enumitem}
\usepackage{amsmath}
\usepackage{graphicx}
\usepackage{hyperref}
\usepackage[utf8]{inputenc}
\usepackage{CJKutf8}
\usepackage[ruled,linesnumbered,vlined]{algorithm2e}
\setcopyright{rightsretained}

\acmDOI{10.475/123_4}

\acmISBN{123-4567-24-567/08/06}

\acmConference[2018]{ABC}{Jan 2019}{ZZY}
\acmYear{1997}
\copyrightyear{2016}

\acmArticle{4}
\acmPrice{15.00}

\begin{document}
\title{Answering Analytical Queries on Text Data with Temporal Term Histograms}

\author{Kai Lin}
\affiliation{%
  \institution{Univ. of California San Diego}
  \streetaddress{9500 Gilaman Drive}
  \city{La Jolla}
  \state{CA}
  \postcode{92093}
}
\email{klin@sdsc.edu}

\author{Subhasis Dasgupta}
\affiliation{%
  \institution{Univ. of California San Diego}
  \streetaddress{9500 Gilman Drive}
  \city{La Jolla}
  \state{CA}
  \postcode{92093}
}
\email{sudasgupta@ucsd.edu}

\author{Amarnath Gupta}
\affiliation{%
  \institution{Univ. of California San Diego}
  \streetaddress{9500 Gilaman Drive}
  \city{La Jolla}
  \state{CA}
  \postcode{92093}
}
\email{a1gupta@ucsd.edu}

\renewcommand{\shortauthors}{K. Lin et al.}

\begin{abstract}
Temporal text, i.e., time-stamped text data are found abundantly in a variety of data sources like newspapers, blogs and social media posts. While today's data management systems provide facilities for searching full-text data, they do not provide any simple primitives for performing analytical operations with text. This paper proposes the \textit{temporal term histograms}(TTH) as an intermediate primitive that can be used for analytical tasks. We propose an algebra, with operators and equivalence rules for TTH and present a reference implementation on a relational database system.
\end{abstract}
%
%



\maketitle

\section{Introduction}
\label{sec:intro}
Today's data management systems are designed to support data heterogeneity and analytical operations\cite{8258260,6758829}. In particular, all DBMSs today have support for full-text search for text-valued columns, and more complex aggregate operations. For example, PostgreSQL has two data types for full-text search - \textit{tsvector} represents a document as a vector of terms, and \textit{tsquery} represents a predicate over search terms where terms are combined by Boolean operators and a \textit{followedBy} operator to construct phrases. On the analytical side, PostgreSQL supports a number of statistical operations, window operations, temporal operations of different granularities, sampling operations and so on. Other DBMSs like DB2, Oracle, Vertica, SQL Server have comparable set of facilities.

But how well do today's DBMSs support text analytics, i.e., the need to perform analytical operations on textual data?

A number of application domains today, including computational social science applications are shifting to a paradigm called \textit{Text as Data} \cite{grimmer2013text,gentzkow2017text} whose goal is to use construct statistical models of diverse text-centric documents like financial news, legislative text, social media content and company filings, and perform statistical inferences based on these models\cite{DBLP:conf/cidr/SiddiquiLKXYZGL17, LytrasRD17}. When the textual data is time-stamped, temporal analytical tasks include finding significant temporal trends in document features, or comparison between subsets of documents from two temporal intervals are similar. 

\noindent \textbf{Running Example} Suppose, we have a relation of newspapers with the following schema: 
\begin{verbatim}
articles(id:int, newspaper:string, title:string, 
section:string, city:string, pubDate:date, content:text)
\end{verbatim}
Our target user is a an analyst from the domain of social sciences. Let us consider the following analytical questions -- we consider a question to be ``analytical'' if it needs any complex computation beyond data retrieval and standard aggregate functions provided by modern data management platforms. 
\begin{description}[leftmargin=*]
\item[Q1. Topic Co-occurrence.] \textit{For all $n$-day intervals where "tax cut" was a dominant topic (i.e., within the top $k$ key terms), what other topics were also dominant? How does this vary over cities?} 
\item[Q2. Salience Detection] \textit{Given a term set $\mathbf{T}$ is there a week  $i$ such that the set $\mathbf{T}$ is \textbf{salient} for that week?} The set $\mathbf{T}$ is \textit{\textbf{salient}} for a week if all terms in $\mathbf{T}$ are in the top $k$ terms of that week and their rank distribution is significantly different with respect to every other week. 
\item[Q3.Synchronized Topics.] \textit{Which newspapers show perfectly time-coordinated (same top $k$ terms on the same day) articles in the politics section?} This class of queries are used for detecting propaganda and censorship  in state-sponsored newspapers \cite{roberts2018censored}.
\item[Q4. Trendy Topics.] \textit{Find documents dated \texttt{today()-6 months} that contain  ``trendy'' topics.} where terms are ``trendy'' in a time interval  if they  have a count above the average over all terms for that interval, and have a positive temporal gradient greater than a threshold $\theta$. 
\end{description}
Each of these questions hinges on the interplay between the temporal dimensions of the data, the frequency distributions of terms of the \texttt{content} field, and their co-variation with other attributes like \texttt{city} and \texttt{section}. However, they are hard to write as single queries and would require a complex database program requiring deep expertise in database technology, which might be beyond the expertise of our target user group. 

\noindent \textbf{Larger Context and Challenge.} It is now recognized that supporting analytic workloads over relational data needs an analytics friendly data algebra that can express a wide range of analytical operations needed, for example, for machine learning tasks. Linear algebra has been proposed as the new algebra for these workloads, and mappings between linear algebra and relational algebra are being defined. However, we are seeking an implementable intermediate abstraction that can support linear algebra like vector and matrix operations, and yet is implementable over today's technology platforms. The representation we seek must be simple enough to be usable by our target users and yet be expressive enough to serve as a bridge to linear algebra processors. To address this gap, we propose an abstract data type, called the \texttt{temporal term histogram} as a database primitive that can be used at a higher-level than current database abstractions of text. We argue that this primitive captures a core abstraction of many computations underlying the \textit{Text as Data} analysis paradigm.

\noindent \textbf{Our Contributions.} The contributions of this paper are as follows.
\begin{itemize}[leftmargin=*]
\item We present a new data model and algebra for temporal term histogram
\item We present a number of query optimization rules over the above algebra
\item We present an implementation scheme for temporal term histograms over a relational DBMS
\item We present a number of experiments to demonstrate the factors that influence the query performance of the above algebra
\end{itemize}

\section{The Temporal Time Histogram}
\label{sec:tthi}
\begin{figure*}[t] 
\begin{minipage}[t]{0.33 \textwidth}
\begin{tabular}{c c l} \small
\textbf{doc.id} &\textbf{doc.ts} &\textbf{doc.text}\\
1 &1 &A B C B\\
2 &1 &D C A A \\
3 &2 &A E D B\\
\end{tabular}
\end{minipage}
\begin{minipage}{0.33  \textwidth}
\begin{tabular}{c c l} \small
\textbf{doc.id} &\textbf{doc.ts} &\textbf{doc.hist}\\
1 &1 &A 1 \\
1  &1  &B 2\\
1  &2  &C 1\\
2 &1 &A 2\\
2  &1  &C 1\\
2  &1  &D 1 \\
3 &2 &A 1\\
3  &2  &B 1\\
3  &2  &D 1\\
3  &2  &E 1\\
\end{tabular}
\end{minipage}
\begin{minipage}{0.33 \textwidth}
\begin{tabular}{c c c l} \small
\textbf{term} &\textbf{doc.ts} &\textbf{doc.count} &\textbf{doc.ids}\\
A &1 &2 &[1, 2]\\
A &2 &1 &[3]\\
B &1 &2 &[1] \\
B &2 &1 &[3] \\
C &1 &2 &[1, 2]\\
C &2 &0 &[]\\
D &1 &1 &[1] \\
D &2 &1 &[3] \\
E &1 &0 &[]\\
E &2 &1 &[3]\\
\end{tabular}
\end{minipage}
\caption{Given three documents over a vocabulary of 5 terms (left), the MTTH (right) merges the rows of the document histogram (middle), grouping them by term ID and timestamp. If auxiliary variables are present (not shown here), they will also be included in the group-by operation.}     
\label{fig:tth}
\end{figure*}
\subsection{The Model}
\label{sec:model}
Let $\mathcal{C}$ be a collection of documents. A document $d$ minimally has an ID denoted by $d.id$, a timestamp of creation denoted by $d.ts$, and textual content denoted by $d.text$, which can in turn be considered a collection of terms, $d.text.terms$. 

Given the above minimal setting, our first component model is based on term distributions. For simplicity, we do not go into the definition of what makes a term. In practice, issues like whether stop words belong to terms, words after stemming should be considered a single term, or multi-word phrases should be considered a single term etc. do not affect the model and are left for the application. The collection of unique terms over all documents in $\mathcal{C}$ is called $V(\mathcal{C})$, the \textit{Vocabulary} of $\mathcal{C}$. In a vocabulary, every term $t$ is represented by a unique ID $t.id$. Henceforth, we will use the term ID to be the proxy for a term.

A \textit{document histogram} is a tuple $h(d.id, d.ts, d.hist)$ where $d.ts$ is a discretized time interval (e.g. per day) and $d.hist$ is a 2-column matrix -- the first column has the term ID and the second column has the count of the term in the document. 

\begin{table*}[h] \small
\caption{Algebraic Operators over the Vocabulary and the Temporal Term Histogram Index. In this table, we omit some straightforward operations like \textit{getRecords}, that fetches the records given their tupleIDs, and \textit{sort}.}
\begin{tabular}{|p{2in}|p{2.50in}|p{2in}|} 
\multicolumn{1}{c}{\textbf{Operator}} &\multicolumn{1}{c}{\textbf{Description}} &\multicolumn{1}{c}{\textbf{Use}}\\ \hline 
 \textbf{lookup}: term $\rightarrow$ termID &Given a term returns its ID & Usually the first step of a search \\ \hline
\textbf{project} ($\pi$): [attrib] $\rightarrow$ bindings &The standard relational projection operation on the TTH structure. It returns the bindings on the projected attributes & Find only the terms from the TTH, which may be the result of another operation, typically \textbf{select}. \\ \hline
\textbf{select}: predicate $\rightarrow$ TTH & the predicate is over all four properties of a TTH structure, returns a subset of the TTH structure matching the predicate &Return the temporal histogram of documents where count(`Obamacare') $>$ 3 \& count(`Trump') = 0 \& ts $>$ ('06/01/2017') \& doc.id $\in [201-399]$\\ \hline
\textbf{indexOp}: (Op, Idx1, Idx2) $\rightarrow$ [doc.id] &Op can be intersection, difference or union of two document index lists.   & The exploration process may want to expand the result set for a query by adding the results of another query -- this needs a union of indexes. \\ \hline
\textbf{queryIndex}: ([qDocID], timeStart? timeEnd?) $\rightarrow$ [termID, ts] &Given qDocID, a list of query document IDs, returns a set of term-timestamp combinations where the terms occur the query documents. Optionally, the operation is restricted to a specified time interval   & Which terms are included in documents [4000-8000], provided the term document combination occurred in 2016?\\ \hline
\textbf{coarsen}: (newWidth, timeStart?, timeEnd?) $\rightarrow$ TTH &The operation reduces the time granularity of an existing TTH to the specified newWidth, which must be a multiple of the current width. The count value will be recomputed based on the newWidth. Optionally, this operation would be carried out only on a specified time interval marked by the start and/or end parameters. & Return the temporal term histogram aggregating data by the week instead of by the day \\ \hline
\textbf{merge}: (TTH$_1$, TTH$_2$) $\rightarrow$ TTH &The operation merges two TTHs that are temporally aligned. The merged histogram has terms from both TTHs and adds the count of the common term, while performing a union of their doc.Idx &This operation will be used to merge the TTHs obtained from two different queries on the same base data. \\ \hline
\textbf{group}: (TTH, $<$grouping-vars$>$) $\rightarrow$ $<$group-value$>$[TTH]&The operation creates an array of TTHs indexed by values of the grouping variables. The size of the array is that of the Catersian product of the active domains of the grouping variables. &Grouping will mostly be used for the auxiliary attributes.  \\ \hline
\textbf{apply}: function($<$parameters$>$) $\rightarrow$ [value] &The \textbf{apply} operation calls a histogram function like \texttt{min, max, findModes, findMoments(k)}. The nature of the value depends on the function called. The function can be multi-valued e.g., modes of the histogram. & Apply can be used to find peaks and troughs in the 3D histogram\\ \hline
\textbf{applyArg}: function($<$parameters$>$) $\rightarrow$ [(value, [recordID])] &This is a convenient variant of the \textbf{apply} function that returns the result of the function together with set of records associated with each returned value (when the function is multivalued) & Find time intervals where there is burst of terms and identify the corresponding terms \\ \hline
\textbf{sortByAxis}: (axis) $\rightarrow$ TTH &The operation on a single TTH and rearranges it by sorting it by the \texttt{axis} parameter which can either be \texttt{term} or \texttt{count}. The time axis is not impacted by the operation & \\ \hline
\textbf{distance}: (TTH$_1$, TTH$_2$, distance-function) $\rightarrow$ float &Given two TTH structure, the operation applies the distance-function to the frequency distributions and returns numerical measure of their distance & Depending on the application, the distance-function can be simple as a Euclidean distance or complex as a KL divergence \\ \hline
\textbf{collapse:} (axis) $\rightarrow$ 1D histogram &In this operation, the axis can be \texttt{term} or \texttt{ts} -- it produces a marginal histogram over only the time axis (i.e., over all terms) or over the term axis. The doc.id list gets recomputed as part of the operation.  &When time axis is collapsed, the result can be used to construct a term-document matrix that can be used to compute topic models.  \\ \hline
\textbf{extractAxis}: (axis) $\rightarrow$ Vector(axis-Value) &Given an axis (i.e., \texttt{term} or \texttt{ts}), it extracts the values of the axis as a vector   & Given a selection query which produces a reduced TTH, this operation may be used to get the term vector corresponding to the result.  \\ \hline
\end{tabular}
\label{tab:TTHOps}
\end{table*}
The \textit{Minimal Temporal Term Histogram} (MTTH) can be viewed as a 4-tuple 
$MTTH(term.ID, ts, count, list(doc.id))$ created by merging of multiple document histograms as shown in Fig. \ref{fig:tth}, where $ts$ is a discretized time interval as before and $count$ is the term count over all documents within the specific interval. The start and width of the time intervals are specified by the user during setup time, and are part of the configuration parameters of the model. 

The MTTH can be slightly extended by introducing a set of auxiliary attributes $c_1, c_2 \ldots c_k$. For the present paper, we will consider only attributes with discrete domains like \texttt{city, newspaper} and \texttt{session} in our running example. The extended model called the \textit{Temporal Term Histogram} (TTH) is a $4+k$-tuple. If $adom(i)$ is the active domain of auxiliary attribute $c_i$ of TTH, each row of the projection $\pi_{c_1, c_2, \ldots c_k}(TTH)$ is unique. In other words, every combination of values from $adom(1) \times \ldots \times adom(k)$ is represented only once in the projection.
\subsection{Operators}
\label{sec:ops}
One can view a temporal term histogram as a hybrid structure between an inverted index and a pure histogram model, because it uses both the vocabulary $V$ and $TTH$; the algebraic operators for the model are listed in Table \ref{tab:TTHOps}. In this model we assume that every tuple of the TTH structure has a unique tupleID. In the following, we point out some salient features of this operator set.
\begin{itemize} [leftmargin=*]
\item The TTH is a mostly a relational structure except for the docID list. So it admits extractive relational operations like \textbf{select} ($\sigma$) and \textbf{project} ($\pi$). The union operation is replaced by \textbf{merge}, which requires the count variable to be added if the (term, ts) pair matches, and behave like a union otherwise (i.e., a \textit{bag union} semantics). Similarly, the docIDs would be merged for the matching pairs. For example, the merge operation of two TTHs is shown in Figure \ref{fig:tth-merge}.
\begin{figure*}[t] 
\begin{minipage}[t]{0.33 \textwidth}
\begin{tabular}{c c c l} \small
\textbf{term} &\textbf{doc.ts} &\textbf{doc.count} &\textbf{doc.ids}\\
A &1 &2 &[1, 2]\\
A &2 &3 &[1, 3]\\
B &1 &4 &[2, 3, 4] \\
B &2 &1 &[4] \\
\end{tabular}
\end{minipage}
\begin{minipage}[t]{0.33 \textwidth}
\begin{tabular}{c c c l} \small
\textbf{term} &\textbf{doc.ts} &\textbf{doc.count} &\textbf{doc.ids}\\
A &1 &3 &[5, 6]\\
B &1 &2 &[5] \\
B &2 &1 &[6]\\
C &2 &1 &[5] \\
\end{tabular}
\end{minipage}
\begin{minipage}[t]{0.33 \textwidth}
\begin{tabular}{c c c l} \small
\textbf{term} &\textbf{doc.ts} &\textbf{doc.count} &\textbf{doc.ids}\\
A &1 &5 &[1, 2, 5, 6]\\
A &2 &3 &[1, 3]\\
B &1 &6 &[2, 3, 4, 5]\\
B &2 &2 &[4, 6] \\
C &2 &1 &[5] \\
\end{tabular}
\end{minipage}
\caption{The merge operation for the left two TTH objects results in the third TTH object on the right. A new term adds a new row while merging on an existing term for the same time, increases the count and extends the document list.}     
\label{fig:tth-merge}
\end{figure*}
\item Since TTH can be viewed as a histogram over terms and time, the operators offer a number of histogram manipulation operations. Specifically, it permits \textit{axis manipulation} using \textbf{coarsen}, \textbf{sortByAxis} and \textbf{collapse} operations as well as builtin and user-defined aggregation operations that can be invoked using the \textbf{apply} operation. Operation \textbf{applyArg} is motivated by the commonly used \textit{argmax} and \textit{argmin} functions, and finds not only the value of the function (e.g., the peaks and troughs of the histogram) but also identifies the records (i.e., effectively the term, time and documents) for which this value occurs. 
\item Since the TTH is an inverted index, it allows index operations that are typically used for answering Boolean queries. Standard operations for \textbf{list intersection} and \textbf{list union} are enabled by the \textbf{IndexOp} operation. The \textbf{queryIndex} operation accepts a list of document IDs and finds term-time pairs in those documents. We recognize that like any inverted index, TTH does not obviate the need for a document-wise forward index. Consider the query "Find the term frequency of the term `Trump' in all documents where `Obama' also appears". The query would first fetch documents where both terms occur (e.g., by index intersection); then it would need to find the frequency of the term `Trump' in the documents from the prior result. However, the TTH does not maintain a forward index from a document to the terms it contains. Thus, the system will need a separate forward index to answer this query properly.
\item A primary use of the auxiliary attributes in the TTH is that they are primarily used as grouping variables. The \textbf{group}($\gamma$) operation in the TTH algebra is used for simply constructing the groups, and a partitioned relation over the TTH tuples for each distinct value of the grouping variable(s) -- no aggregation operations are associated with grouping. A partition is identified by the specific value of the grouping variable it corresponds to. Thus, if the \texttt{city} variable has three values \texttt{NY, LA, SFO} the result of $\gamma_{city}(TTH)$ will produce $TTH_{NY}, TTH_{LA}$ and $TTH_{SFO}$ respectively. As we show in Section \ref{sec:example}, since the group operation can be applied in the middle of a query plan, other operators are designed to operate on partitioned TTH arrays. 
\end{itemize}

TTH enables us to express some analytical queries that cannot be expressed through a full-text query enabled DBMS like PostgreSQL or standard inverted index server like Apache Solr and can be complex to formulated in RDBMS. 

\subsection{Query Examples}
\label{sec:example}
In Section \ref{sec:intro}, we introduced a set of analytical questions. Here we present query plans for these questions based on the algebraic operations in Table \ref{tab:TTHOps}, broken up into phases. Our goal in this section is to write the query plans in terms of TTH and standard relational operations. These plans are not optimized. We assume that the default temporal granularity of TTH is 1 day.

\noindent \textbf{Q1. Topic Co-occurrence.} Since the default temporal granularity is 1 day and the desired granularity is 5 days, we first apply the \textbf{coarsen} operation. Then we filter non the term ``tax cut'' and sort the results on the count field in descending order, and pick the top 20 records. From these records, we project the timestamp field to find the times when the selected term is most used
\begin{multline*}
\pi_{ts}(\sigma_{top(20)}(sort_{count,d}\\
(\sigma_{term='tax cut'}(coarsen('5 days')))))(TTH)
\end{multline*}
where the $d$ parameter performs a sort in descending order.

\noindent Next, we select the top 21 distinct terms from TTH in the same time period as above and then remove the term ``tax cut'' from the result.
\begin{multline*}
\pi_{term}(\sigma^D_{top(21)}(sort_{count,d}(\sigma_{ts~ \in~ \bullet}\\
(coarsen('5 days'))))(TTH) - term('tax cut')
\end{multline*}
where $\bullet$ indicates the result of the first step and $\sigma^D$ refers to a select distinct operation.

\noindent The second part of the query (``how does it vary by city''), we perform the above two steps except that we add a \textbf{group} operation on the variable \texttt{city} as follows.
\begin{multline*}
\pi_{ts}(\sigma_{top(20)}(sort_{count,d}(\gamma_{city}\\
(\sigma_{term='tax cut'}(coarsen('5 days'))))))(TTH)
\end{multline*}
The result of the query will produce a set of timestamps per city.
\begin{multline*}
\pi_{term}(\sigma^D_{top(21)}(sort_{count,d}((\sigma_{ts~ \in~ \bullet}\\
\gamma_{city}(coarsen('5 days')))))(TTH) - term('tax cut')
\end{multline*}
The innermost grouping creates a city-based partition over the coarsened data. The selection operation over timestamp and every upstream operation is performed per partition, leading to an set of terms per city.

\noindent \textbf{Q2. Salience Detection.} Since the query is on weekly term ranking, one way to formulate the query is create a \textit{TTH view} that transforms the default count-based TTH to a rank-based TTH, called TTH$_R$ here. For practical purposes, the view would be computed only on weekly aggregate data for which the \texttt{doc.count} exceeds a given threshold $ct$, and the \texttt{doc.list} column is omitted. We first write the view definition for TTH$_R$.
\begin{multline*}
\pi_{term,ts,\rho(rank(), rank)}(sort_{count,d}\\
(\sigma_{doc.count > ct \wedge all(terms \in T)}(coarsen('7 days'))))(TTH)
\end{multline*}
where $T$ is the list of terms given by the user, the rank() function reports the sort order of a tuple, and $\rho$ is the rename operation. Further, the predicate $all(terms \in T)$ is universally quantified, i.e. the predicate only selects weeks in which all terms appear in some document.

\noindent Now we need to choose a week for which the aggregate rank of the terms in $T$ is the highest using any suitable aggregate function. We use the sum of their ranks $R_{sum}$ as an example. Once we have the $R_{sum}$ values for the highest-rated week: 
\begin{multline*}
\pi_{ts,sum}(max(sum(group_{ts}(\sigma_{term \in T},rank))))(TTH_R)
\end{multline*}
With the weekly $R_{sum}$ values, we can directly use it for a statistical test like the Mann Whitney U Test \cite{hollander2013nonparametric}, where the $R_{sum}$ is directly compared with that of every other week. 
\noindent \textbf{Q3. Synchronized Topics.}

\noindent \textbf{Q4. Trendy Terms.} we can formulate the query in the following steps.
\begin{enumerate}[label=\alph*), leftmargin=*]
\item select a part of the TTH for the specified time period and whose count value is greater than AvgCount, which is the average of the count of all terms in the same time period, computed separately. 
\vspace{-2mm}
$$\sigma_{ts > today() - months(6) ~\wedge~ count > AvgCount}(TTH)$$
\item Next, we would apply a function that extracts the maximum slope from the count value along the time axis and identify the corresponding records. The $\bullet$ designates the partial result from the previous step. Select the results where the value part of the result exceeds the threshold.
\vspace{-2mm}
$$\sigma_{value > \theta}(ApplyArg(findMaxSlope(count,ts), \bullet))$$
where the user-defined findMaxSlope function accepts the axis $ts$ and the variable $count$ over which the function would be computed.
\item Project the record IDs (2nd variable) from the results of the filter, and extract their docIDs. Then, find the docIDs for these records.
\vspace{-2mm}
$$\pi_{doc_id}(getRecords(\pi_2(\bullet)))$$
\end{enumerate}
If in the process of exploration we want to apply the findMaxSlope function on the time granularity of a week instead of a day, we will use the \textbf{coarsen} operator. In that case, the second segment of the query will be modified to:
\vspace{-2mm}
$$\sigma_{value > \theta}(ApplyArg(findMaxSlope(count,ts), coarsen(\bullet)))$$
\vspace{-2.5mm}

\noindent \textbf{Algebraic Optimization.} Many mathematical properties are satisfied by Term Temporal Histogram operations.  The following are some example properties between the coarsen and merge operations:
\begin{enumerate}[leftmargin=*]
\item The associativity of the merge operation:
$$  tthi\_merge(tthi\_merge(X, Y), Z) 
= tthi\_merge(X, tthi\_merge(Y, Z)) $$

\item The commutativity between the merge and coarsen operation:     \begin{align*}tthi\_coarsen(tthi\_merge(X, Y),  D) \\
= tthi\_merge(tthi\_coarsen(X, D), tthi\_coarsen(Y,D))
\end{align*}
\item The quasi-idempotence of the coarsen operation:
$$tthi\_coarsen(tthi\_coarsen(X,  D1),  D2)
=  tthi\_coarsen(X, D2)$$
\end{enumerate}
Using these properties can optimize a complicated histogram expression.  For example, the left hand side of the commutativity above requires merging two very large histograms X and Y, which could be very slow, and then convert the merging result into a new histogram with larger data ranges; however, the right hand side first converts two large histograms into two smaller histograms and then merge them.  Using the expression of the right hand side to calculate the left hand side could have better performance.
\section{A Reference Relational Implementation}
\label{sec:tth-processor}
We have implemented the TTH model on top of a relational database (in our case, PostgreSQL). We have assumed that the documents in question are text-valued attributes of relations. A relation may have multiple text attributes to be indexed by the TTHI structure. In the following, we outline the steps to construct and use a temporal term histogram.

\noindent \textbf{Mapping Declaration.} To construct a Temporal Term Histogram from a relational schema, the user must first identify the schema elements within the purview of the index. This is performed through a set of annotated mapping statements.  Consider the schema
\begin{verbatim}
newspapers(name, province, publisher)
newsArticle(id, newspaper, title, date, author, page, 
            content)
\end{verbatim}
where the data is extracted from CSV files, the following mapping statements would identify text and temporal attributes in \texttt{newsArticle}.
\begin{verbatim}
@CSV(corpus = "chinese-newspaper", entity = "Newspaper")
   @CSVColumn(mapto = "date")
   @Temporal(TemporalType.DATE, format="YYYY-MM-DD")
    Date date;
   @TermIndex(mapto = "title", language="Chinese")
    String title;
   @TermIndex(mapto = "content", language="Chinese")
    String content;
\end{verbatim}
The first statement declares that a corpus called ``chinese-newspaper'' is a CSV file and is henceforth called by the name ``Newspaper''. By default all attributes actually used in constructing the TTHI are referred to as \texttt{CSVColumn}. Since the attribute \texttt{newsArticle.author} is not relevant for the TTH, it is not captured in any mapping statement. Some of the columns have a special role in the TTH. An attribute annotated as \texttt{TermIndex} with data type \texttt{String} is interpreted as a text field that will be indexed on TTH construction algorithm. Similarly, the attribute that serves as the time indicator is annotated as \texttt{@Temporal}. If the TTH is constructed over multiple tables, one can also specify foreign keys between tables. For example, the reference from \texttt{newsArticle.newspaper} to \texttt{newspapers.name} is stated as:
\begin{verbatim}
@CSVColumn(mapto = "newspaper.name")
    String newspaper;
\end{verbatim}
The mapping specification also allows the specification of \textit{category attributes}, attributes for which we would like to create additional axes for specifying the histogram. To declare \texttt{newspaper.province} as a category, we state:
\begin{verbatim}
@Category(name=province)
   String province;
\end{verbatim}
In this case, the term count for the histogram will be based on for each value of the cross product of \texttt{date}, \texttt{term}(from \texttt{content}, and separately, from \texttt{title}) and \texttt{province}.

\noindent \textbf{Relational Implementation.} Given the mapping declaration, the system constructs a set of base tables that store the histogram for each \texttt{TermIndex} column for each time-unit (day in the above specification) and category column. Algorithm \ref{algo:createTables} shows the process of constructing these tables.

\begin{algorithm}
\For{each entity \textbf{ent} in the corpus}{
   create a table \textbf{tab} with the specified name or the entity name\;
   
   \For {each attribute \textbf{att} of \textbf{ent}} {
  	   \eIf{ the type of \textbf{att} is an entity \textbf{ent'}}{
   		   create a column in \textbf{tab} with the name of \textbf{att} and 
   		   the postfix \textsf{\underline{ }id} and with the type of the 
   		   \textsf{@Id} attribute of \textbf{ent'}\;
   	   }{
           create a column in \textbf{tab} with the name and correspondent type 
           of \textbf{att}\; 
       }       
    }
    \If {\textbf{ent} is the main entity of the corpus} {
       \For {each term index \textbf{tid}}{
       	   create a \textsf{tsvector} column with the name of \textbf{tid} and 
       	   \textsf{\underline{ }tsv}
       }
    }
}

\For{each term index \textbf{tid}}{
   create a table \textbf{tab} with the name of \textbf{tid} and the postfix
   \textsf{\underline{ }doc\underline{ }frequency}\;
   
   \For {each attribute of \textbf{tid}} {
  	   create a \textsf{string} column in \textbf{tab} with the name of \textbf{tid}\;
   } 
   
   create a column \textsf{doc\underline{ }id} in \textbf{tab} as the foreign key of 
   the main entity\;
   
   create an \textsf{int} column count in \textbf{tab}\;
}

\For{each term index \textbf{tid}}{
   \For {each category of \textbf{cat}} {
  	   create a table \textbf{tab} using the concatenation of the	 name of  	 
  	   \textbf{tid}, \textbf{cat} and the postfix 
  	   \textsf{\underline{ }base\underline{ }histogram} as the table name\;
  	   
  	   create a \textsf{string} column in \textbf{tab} with the name of \textbf{tid}\;
  	   
  	   create a \textsf{date} column in \textbf{tab} with the name \textsf{date}\;
  	   \For {each attribute \textbf{att} of \textbf{cat}} {
  	      create a \textsf{string} column in \textbf{tab} with the name of \textbf{att}\;
       } 
       create an \textsf{int} column in \textbf{tab} with the \textbf{tid} name and
       \textsf{\underline{ }term\underline{ }count} as the column name\;
       
       create an \textsf{int} column in \textbf{tab} with the name 
       \textsf{doc\underline{ }count} \;
   } 
}

\caption{Create Database Tables for A Corpus (PostgreSQL version)}
\label{algo:createTables}
\end{algorithm}
Thus, for our example case, Algorithm \ref{algo:createTables} produces the following.
\begin{enumerate}[leftmargin=*]
\item For any combination of document attributes that are considered as a histogram object, a new column with the type \textit{tsvector} is added into the document table and a trigger is setup to save the tokens from the columns of these attributes into the \textit{tsvector} column whenever an insert event occurs on the document table.
\item A table \textit{term\_doc\_frequency} is created to record the term frequency in each document for the selected terms in the study.  A trigger is attached to the update event on the tsvector column of the document table to calculate the term frequencies in the tsvector column using the built-in function \textit{ts\_stat} and save the result into the \textit{term\_doc\_frequency} table.
\item A query-specific \textit{histogram} table is generated on the fly as a materialized view of a query against the document table, the \textit{term\_doc\_frequency} table and other metadata tables.  For example, a user can create ``a monthly histogram for the newspaper articles that are published on Beijing Daily in 2017 and contain the term 'Xi Jianping' at least 5 times'' by automatically constructing the view definition in Figure \ref{fig:histogram}. Once the view is defined, subsequent queries can be asked on it. For example, the query
\begin{verbatim}
   SELECT *
     FROM Histogram(NewspaperArticle, 
                    '2017-01-01', '2017-12-31', 10, 
                    province, 
                    content_term)
    WHERE NewspaperArticle.page = '3'
      AND province = 'Beijing'
      AND content_term @@ 'Trump'                
\end{verbatim}
seeks the TTH from news papers with predicated on the auxiliary attributes \texttt{province} and \texttt{page}, and on the term of interest (\texttt{Trump}) for the specified date range. Algorithm \ref{algo:translation} specifies how the above query will be translated to an equivalent query against the histogram created with the materialized view.
\end{enumerate}
\begin{figure}
\begin{verbatim}
CREATE MATERIALIZED TERM_TEMPORAL_HISTOGRAM VIEW 
  beijing_xi_monthly_2017 AS   
  WITH dr AS 
            (SELECT generate_series as date_start, 
                    row_number() OVER () as num
               FROM generate_series('2017-01-01'::date, 
      '2018-01-01'::date, 
      '1 month'::interval)),
       dateranges AS (SELECT daterange(X.date_start::date, 
               Y.date_start::date) 
                 as date_range
                        FROM dr X, dr Y
                       WHERE X.num + 1 = Y.num)
 SELECT term_doc_frequency.term, 
        dateranges.date_range,
        sum(term_doc_frequency.frequency) as count,
        array_agg(term_doc_frequency.doc_id) as doc_list
   FROM newspaper_articles, 
        term_doc_frequency,
	  dateranges
  WHERE newspaper_articles.id = term_doc_frequency.doc_id
    AND newspaper_articles.date::date <@  date_range
    AND newspaper_articles.newspaper = 'Beijing Daily'
    AND term_doc_frequency.doc_id IN
         (SELECT term_doc_frequency.doc_id
            FROM term_doc_frequency
           WHERE term_doc_frequency.term = 'Xi Jianping' 
             AND term_doc_frequency.frequency >= 5)
  GROUP BY term_doc_frequency.term, date_range;
\end{verbatim}
\caption{The system automatically generates materialized views}
\label{fig:histogram}
\end{figure}
The declaration \texttt{MATERIALIZED TERM\_TEMPORAL\_HISTOGRAM VIEW} generates a histogram with columns \texttt{term, date\_range, count} and \texttt{doc\_list},where \texttt{count} gives the frequency of the term in the date\_range; the array \texttt{doc\_list} includes all document ids containing the term in the \texttt{date\_range}. 

The auxiliary query \texttt{dr} generates a list of the first day of each month from 2017-01-01 (included) to 2018-01-01 (excluded).  The query date ranges creates month based data ranges from 2017-01-01 to 2018-01-01 by joining \texttt{dr} with the one month shifted \texttt{dr}.

A typical histogram usually can be created by joining the document table with the table \texttt{term\_doc\_frequency} and the result of the auxiliary query date ranges. In the example above, the table \texttt{newspaper\_articles} is the document table. The condition 
\texttt{newspaper\_articles.date::date <@ date\_range} gets all the documents whose publish dates are in the \texttt{date\_range}, and the predicate on \texttt{newspaper\_articles.newspaper} limits the newspaper to Beijing Daily.  The last condition gets all the documents where the term Xi Jianping occurs at least 5 times.

\begin{algorithm}
\eIf{ conditions\underline{ }for\underline{ }main\underline{ }entry 
	  and conditions\underline{ }for\underline{ }term\underline{ }index are empty}{
	     \eIf{ TimeInterval == 1 }{
	     	 translate the query conditions into a direct query on the base histogram 
	     	 for the category and the term index\;
	     }{
	     	 translate the query conditions into a grouping query on the base histogram 
	     	 for the category and the term index\;
	     }
    }{   
	   translate the query conditions into a grouping query on the join of the main entity 	 	   table and relevant tables and the doc\underline{ }frequency table\;   
    }
\caption{Query Translation}
\label{algo:translation}
\end{algorithm}

\noindent\textbf{Operator Implementation.} We describe the implementation of some of the more complex TTH operations.

\noindent \textit{Coarsen.} The coarsen operation generates a new histogram from an existing histogram by using a larger date range.  For example, the following statement creates a quarterly based histogram for the newspaper articles that are published on Beijing Daily in 2017 and contain the term 'Xi Jianping' at least 5 times:
\begin{verbatim}
SELECT *
  FROM tthi_coarsen(beijing_xi_monthly_2017, 
   '2017-01-01', 
   '2018-01-01', 
   '3 month');
\end{verbatim}
The coarsen operator first \textit{validates} the compatibility of the date ranges of the new histogram with the date ranges in the base histogram. The data range is valid if each date range in the new histogram is aligned with several consecutive date ranges in the base histograms.  The following query is used in the validation for the statement above:
\begin{verbatim}
WITH tmp_tthi_1 AS (SELECT generate_series AS date_start,
        				  row_number() OVER () AS num 
    				   FROM generate_series('2017-01-01'::date,
        							     '2018-01-01'::date,
        							     '3 month'::interval)), 
	tmp_tthi_2 AS (SELECT daterange(X.date_start::date,
                           Y.date_start::date) AS date_range 
                      FROM tmp_tthi_1 X, tmp_tthi_1 Y 
                     WHERE X.num + 1 = Y.num), 
     tmp_tthi_3 AS (SELECT * FROM beijing_xi_monthly_2017) 
SELECT count(*) 
  FROM
      (SELECT DISTINCT date_range FROM tmp_tthi_3) 
            tmp_tthi_4,
      (SELECT DISTINCT date_range FROM tmp_tthi_2) 
            tmp_tthi_5 
 WHERE NOT tmp_tthi_4.date_range = tmp_tthi_5.date_range 
   AND isempty(tmp_tthi_4.date_range * 
               tmp_tthi_5.date_range) = false 
   AND NOT tmp_tthi_4.date_range * 
           tmp_tthi_5.date_range =       
     tmp_tthi_4.date_range
\end{verbatim}
The main idea behind this validation algorithm is that if a date range in the new histogram is overlapping with a date range in the base histogram, then the date range in the base histogram must be a sub-date range of the data range in the new histogram. If the validation fails, the execution of the operation will stop with an error message, otherwise the following query result will be returned as the new histogram:
\begin{verbatim}
SELECT * 
  FROM (WITH tmp_tthi_7 AS 
         (SELECT generate_series AS date_start,
                 row_number() OVER () AS num 
          FROM generate_series('2017-01-01'::date,
            	                '2018-01-01'::date,
            					'3 month'::interval)),
             tmp_tthi_8 AS 
             (SELECT daterange(X.date_start::date,
                               Y.date_start::date) 
					                      AS date_range 
              FROM tmp_tthi_7 X, tmp_tthi_7 Y 
              WHERE X.num + 1 = Y.num),
             tmp_tthi_9 AS (SELECT * 
                              FROM beijing_xi_monthly_2017) 
        SELECT term,
               tmp_tthi_8.date_range,
               sum(tmp_tthi_9.count) AS count,
               array_merge_agg(tmp_tthi_9.doc_list) AS doc_list 
          FROM tmp_tthi_9, tmp_tthi_8 
         WHERE tmp_tthi_9.date_range <@ tmp_tthi_8.date_range 
         GROUP BY term, tmp_tthi_8.date_range 
         ORDER BY term, tmp_tthi_8.date_range) AS tmp_tthi_10
\end{verbatim}
\noindent \textit{Merge.} The merge operation generates a new histogram from two existing base histograms.  For example, the following statement creates new histogram from two existing histograms:
\begin{verbatim}
SELECT * 
  FROM tthi_merge(beijing_xi_monthly_2017, 
       jiefang_xi_monthly_2017);
\end{verbatim}
where jiefang\_xi\_monthly\_2017 is the histogram for the newspaper articles that are published on Jiefang Daily in 2017 and contain the term 'Xi Jianping' at least 5 times. 

The merge operation requires a validation step too before creating the new histogram. The validation uses the following query to check whether the date ranges in the two existing histograms are aligned:
\begin{verbatim}
SELECT count(*) 
  FROM (SELECT DISTINCT date_range 
          FROM beijing_xi_monthly_2017) tmp_tthi_1,
       (SELECT DISTINCT date_range 
          FROM jiefang_xi_monthly_2017) tmp_tthi_2 
 WHERE NOT tmp_tthi_1.date_range = tmp_tthi_2.date_range 
   AND isempty(tmp_tthi_1.date_range * 
               tmp_tthi_2.date_range) = false
\end{verbatim}

The validation makes sure that any different data ranges from the two histograms are not overlapping. If the validation is passed, the following SQL statement will be executed to get the new histogram:
\begin{verbatim}
SELECT * 
  FROM (WITH tmp_tthi_5 AS 
    (SELECT
     CASE WHEN tmp_tthi_3.term IS NOT NULL 
             THEN tmp_tthi_3.term 
          WHEN tmp_tthi_4.term IS NOT NULL 
             THEN tmp_tthi_4.term 
     END AS term,
     CASE WHEN tmp_tthi_3.date_range IS NOT NULL 
             THEN tmp_tthi_3.date_range 
          WHEN tmp_tthi_4.date_range IS NOT NULL 
             THEN tmp_tthi_4.date_range 
     END AS date_range,
     CASE WHEN tmp_tthi_3.count IS NULL   
             THEN tmp_tthi_4.count 
          WHEN tmp_tthi_4.count IS NULL 
             THEN tmp_tthi_3.count 
          ELSE tmp_tthi_3.count + tmp_tthi_4.count 
     END AS count,
     CASE WHEN tmp_tthi_3.doc_list IS NULL 
             THEN tmp_tthi_4.doc_list 
          WHEN tmp_tthi_4.doc_list IS NULL 
             THEN tmp_tthi_3.doc_list 
          ELSE tmp_tthi_3.doc_list | tmp_tthi_4.doc_list 
     END AS doc_list 
     FROM beijing_xi_monthly_2017 AS tmp_tthi_3 
             FULL JOIN jiefang_xi_monthly_2017 AS tmp_tthi_4 
               ON (tmp_tthi_3.term = tmp_tthi_4.term 
              AND tmp_tthi_3.date_range = 
                  tmp_tthi_4.date_range)),
        tmp_tthi_6 AS 
           (SELECT tmp_tthi_3.term,
                   tmp_tthi_3.date_range,
                   tmp_tthi_3.doc_list & 
                   tmp_tthi_4.doc_list 
					      AS doc_list 
            FROM beijing_xi_monthly_2017 AS tmp_tthi_3,
                 jiefang_xi_monthly_2017 AS tmp_tthi_4 
   WHERE tmp_tthi_3.date_range = tmp_tthi_4.date_range 
  AND tmp_tthi_3.term = tmp_tthi_4.term AND array_length
  (tmp_tthi_3.doc_list &
							      tmp_tthi_4.doc_list, 1) > 0),
  tmp_tthi_7 AS (SELECT term, date_range, doc_id 
  FROM tmp_tthi_6, unnest(tmp_tthi_6.doc_list) doc_id),
        tmp_tthi_8 AS 
     (SELECT tmp_tthi_7.term, tmp_tthi_7.date_range,
sum(word_doc_frequency.frequency) AS count,
 array_agg(tmp_tthi_7.doc_id) AS doc_list 
 FROM tmp_tthi_7, word_doc_frequency 
 WHERE tmp_tthi_7.term = word_doc_frequency.word 
AND tmp_tthi_7.doc_id = word_doc_frequency.doc_id 
GROUP BY tmp_tthi_7.term, tmp_tthi_7.date_range) 
SELECT tmp_tthi_5.term, tmp_tthi_5.date_range,
  CASE WHEN tmp_tthi_8.count IS NULL THEN tmp_tthi_5.count 
    WHEN tmp_tthi_8.count IS NOT NULL 
       THEN tmp_tthi_5.count - tmp_tthi_8.count 
     END AS count,
       tmp_tthi_5.doc_list 
 FROM tmp_tthi_5 
        LEFT JOIN tmp_tthi_8 
     ON (tmp_tthi_5.term = tmp_tthi_8.term 
   AND tmp_tthi_5.date_range = tmp_tthi_8.date_range) 
ORDER BY tmp_tthi_5.term, tmp_tthi_5.date_range) 
   AS tmp_tthi_9
\end{verbatim}
where the auxiliary query \texttt{tmp\_tthi\_5} simply adds the term counts and union the document id arrays from the two base histograms for the same term and get the union of the base histograms.  However the two document id arrays from the two base histograms may overlap, i.e., they may contain same document ids in their document arrays and count the term multiple times. The auxiliary query \texttt{tmp\_tthi\_6}, \texttt{tmp\_tthi\_7} and \texttt{tmp\_tthi\_8} calculate the intersection of the base histograms.  The main statement subtracts the intersection from the union to compute the merge result.

\noindent \textit{Histogram Distance.} The TTH implementation allows a number of histogram comparison operations, including a pointwise Euclidean distance. In the following example, the temporal term histograms of for two newspapers are compared based on a monthly granularity. Such a comparison shows the relative importance of terms between the two newspapers.
\begin{verbatim}
SELECT *
FROM tth_eu_distance(beijing_xi_monthly_2017, 
                      jiefang_xi_monthly_2017);
\end{verbatim}

\noindent \textit{TF-IDF Histogram.} The TF-IDF score is normally defined for documents. With a minor abuse of definition, we define the TF-score of a term within a given time interval as the product of the term frequency divided by the log of the proportion of documents containing the term within the time interval. The practical use of this histogram ranking function is to ameliorate the skew of term distribution for terms like (Xi Jianping and Trump Figure \ref{fig:histogram} ) in documents.

For any given term temporal histogram TTH, the operation \texttt{tf\_idf(tth)} returns  a table result with the column rank, term, date range and \texttt{tf\_idf}.  For example, the statement 
\begin{verbatim}
	SELECT * FROM tf_idf(beijing_xi_monthly_2017, 20)
\end{verbatim}
returns the top 20 terms based on the TF-IDF values for each date range (i.e., the results are grouped by monthly intervals) and the IDF score for a term is computed for every group (as shown in Table \ref{tab:TFIDF}). 

\begin{table}
\begin{tabular}{cccc}
rank &term   &date range &tf-idf\\ \hline
1   &\begin{CJK}{UTF8}{gbsn}监督\end{CJK}   &[2017-01-01,2017-02-01) & 0.0193846703149912758\\
2 &\begin{CJK}{UTF8}{gbsn}全球化\end{CJK}   &[2017-01-01,2017-02-01) &0.0179111139022103821\\
3 &\begin{CJK}{UTF8}{gbsn}政协\end{CJK}   &[2017-01-01,2017-02-01) &0.0149677269073624879 \\
4 &\begin{CJK}{UTF8}{gbsn}工作\end{CJK}   &[2017-01-01,2017-02-01) &0.0134366829530191533 \\
5 &\begin{CJK}{UTF8}{gbsn}中央\end{CJK}   &[2017-01-01,2017-02-01) &0.0133800192793237689 \\
6 &\begin{CJK}{UTF8}{gbsn}协商\end{CJK}   &[2017-01-01,2017-02-01) &0.0131263818574010611 \\
$\ldots$ & & &
\end{tabular}
\caption{Result of a TF-IDF operation on a temporal term histogram}
\label{tab:TFIDF}
\end{table}
\begin{figure}
\includegraphics[width=0.47\textwidth]{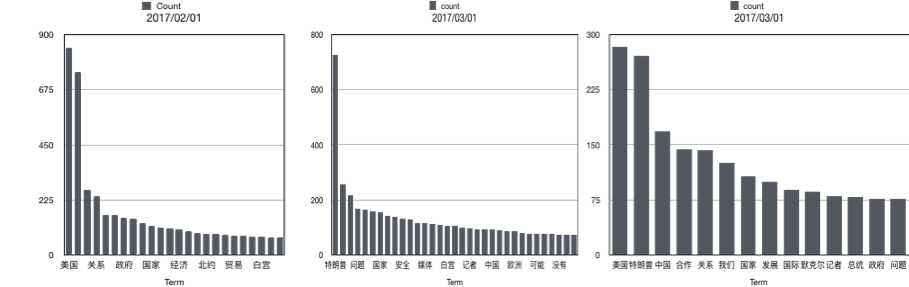}

\caption{Sample Histogram of Trump}
\label{fig:histogram}
\end{figure}

\section{Experimental Evaluation}
\label{sec:experiments}
We used a corpus of 2500 articles from Chinese newspapers containing 6 million distinct terms. The relational implementation of the TTH prototype was created on PostgreSQL 10.1. 

\begin{figure*}
\includegraphics[width=0.47\textwidth]{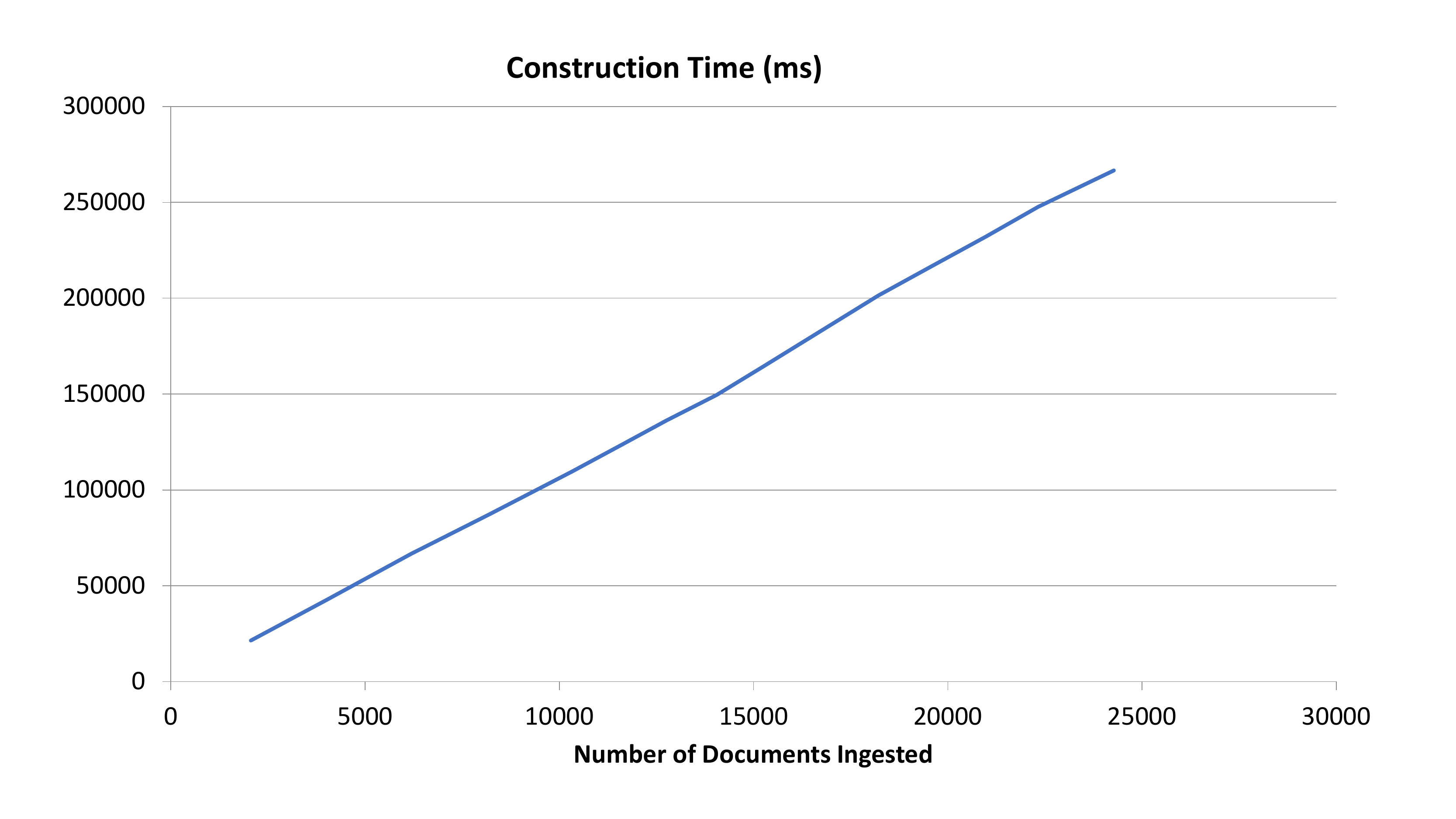}
\includegraphics[width=0.45\textwidth]{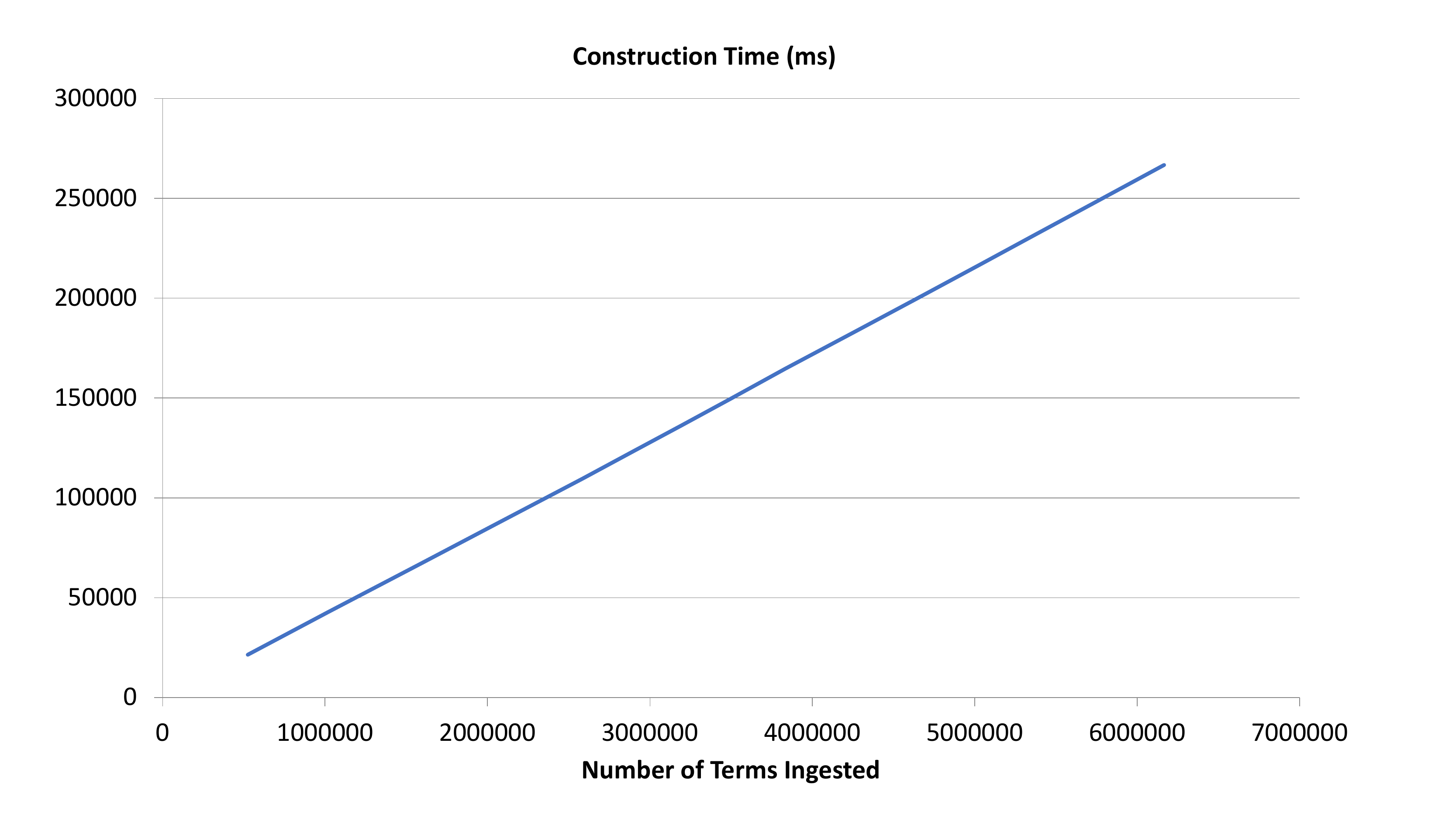}
\caption{The construction of the frequency table is linear with respect to the number of documents and terms}
\label{fig:construction}
\end{figure*}

\noindent \textbf{Construction Cost for Term-Document Frequency Table.} Figure \ref{fig:construction} shows that the time to construct the term document frequency table is linear with document and term sizes. The cost of 300s is high but the Term-Document Frequency table is constructed offline. 

\begin{figure*}
\includegraphics[width=0.47\textwidth]{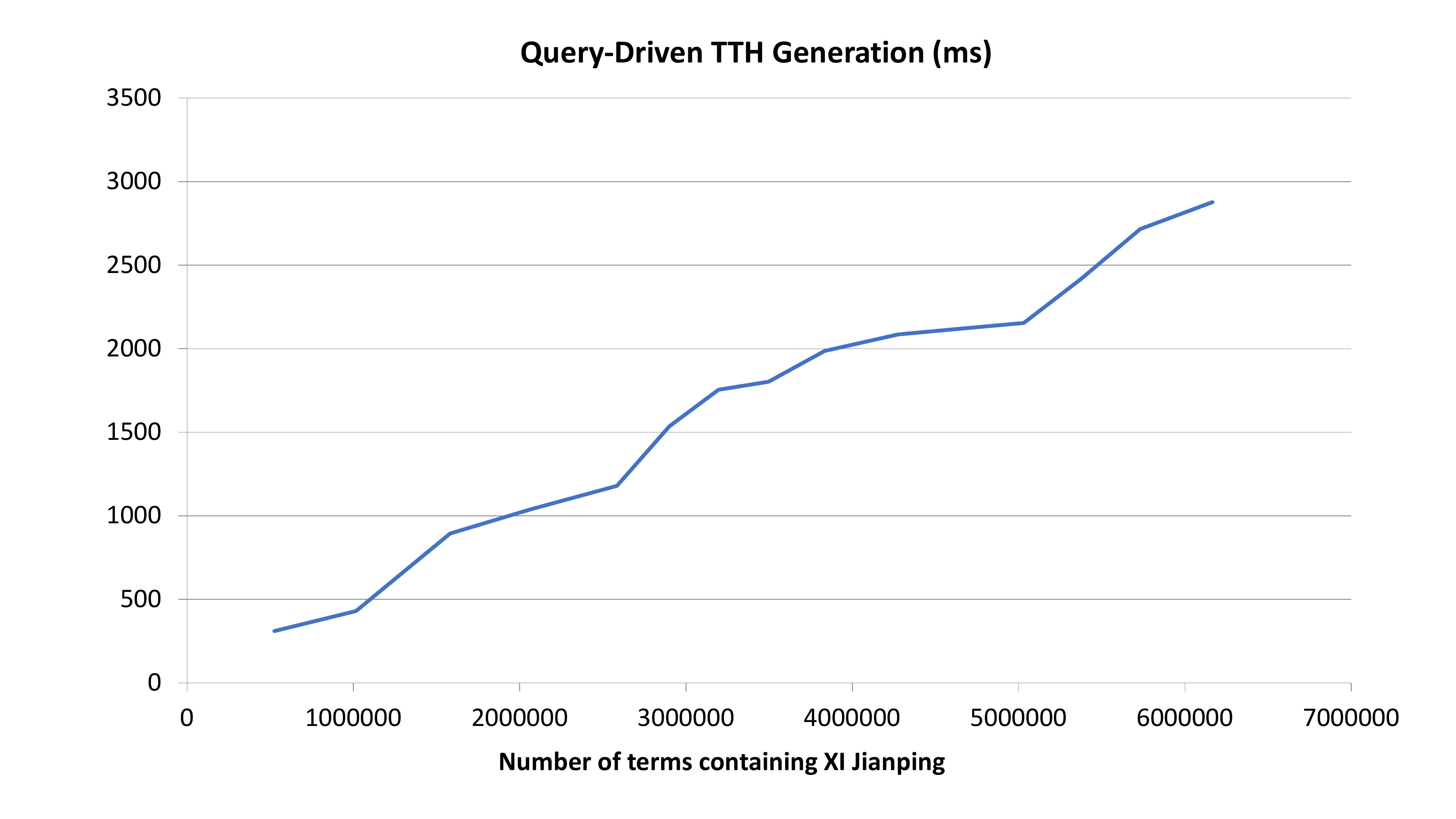}
\includegraphics[width=0.45\textwidth]{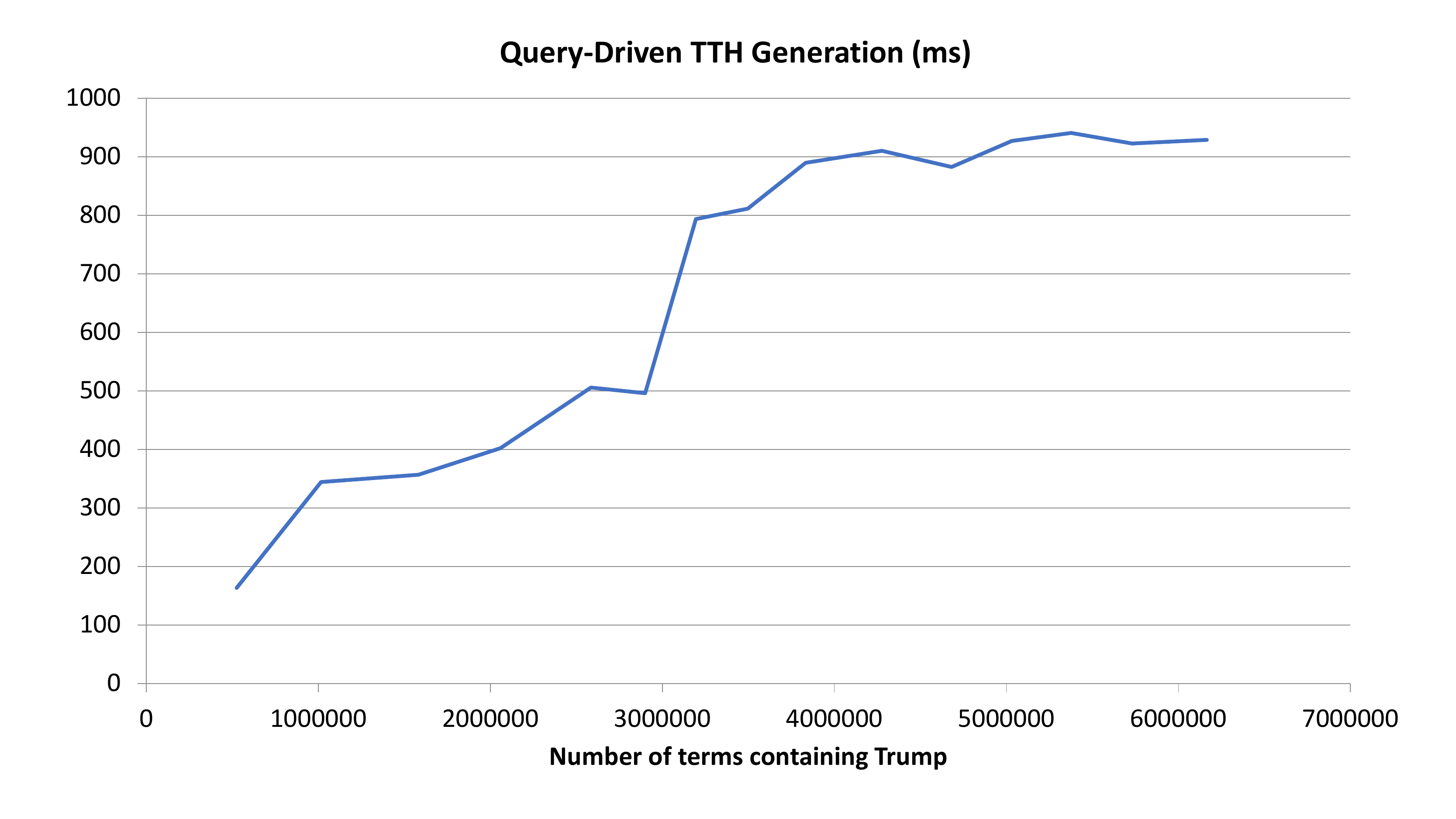}
\caption{The ad hoc TTH creation cost is within interactive response limits. }
\label{fig:adhoc}
\end{figure*}

\noindent \textbf{Ad hoc Histogram Construction Cost.} We generate temporal term histograms on the fly based on user queries. Therefore the cost of histogram construction depends on the number of documents containing the desired terms. Fig. \ref{fig:adhoc} shows the ad hoc TTH creation time for 'Xi Jianping' and 'Trump', the two most popular terms in our corpus. These TTHs are constructed in 3.5s and 1s respectively, which are well within an interactive retrieval limit.

\begin{figure*}
\includegraphics[width=0.47\textwidth]{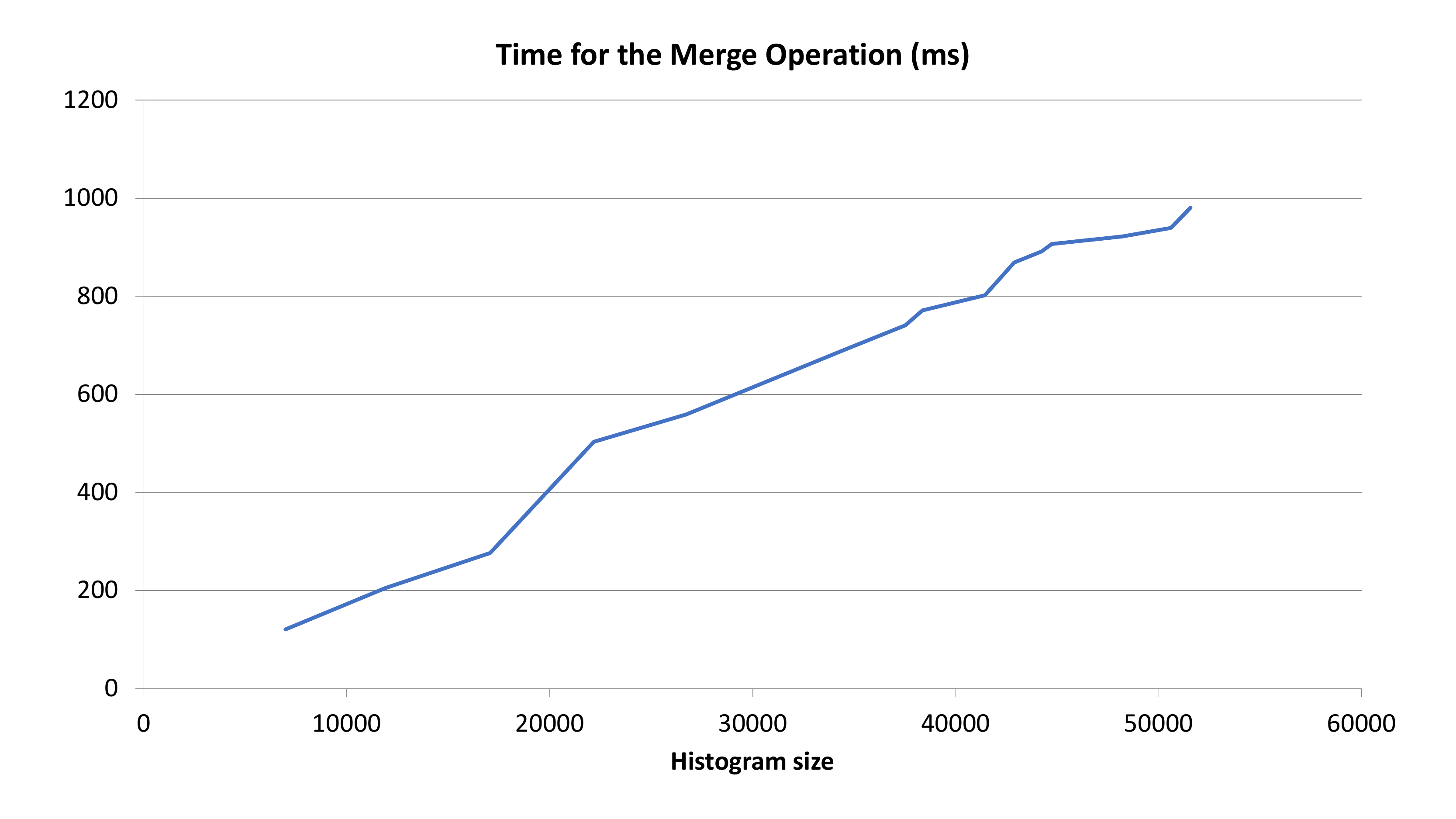}
\includegraphics[width=0.47\textwidth]{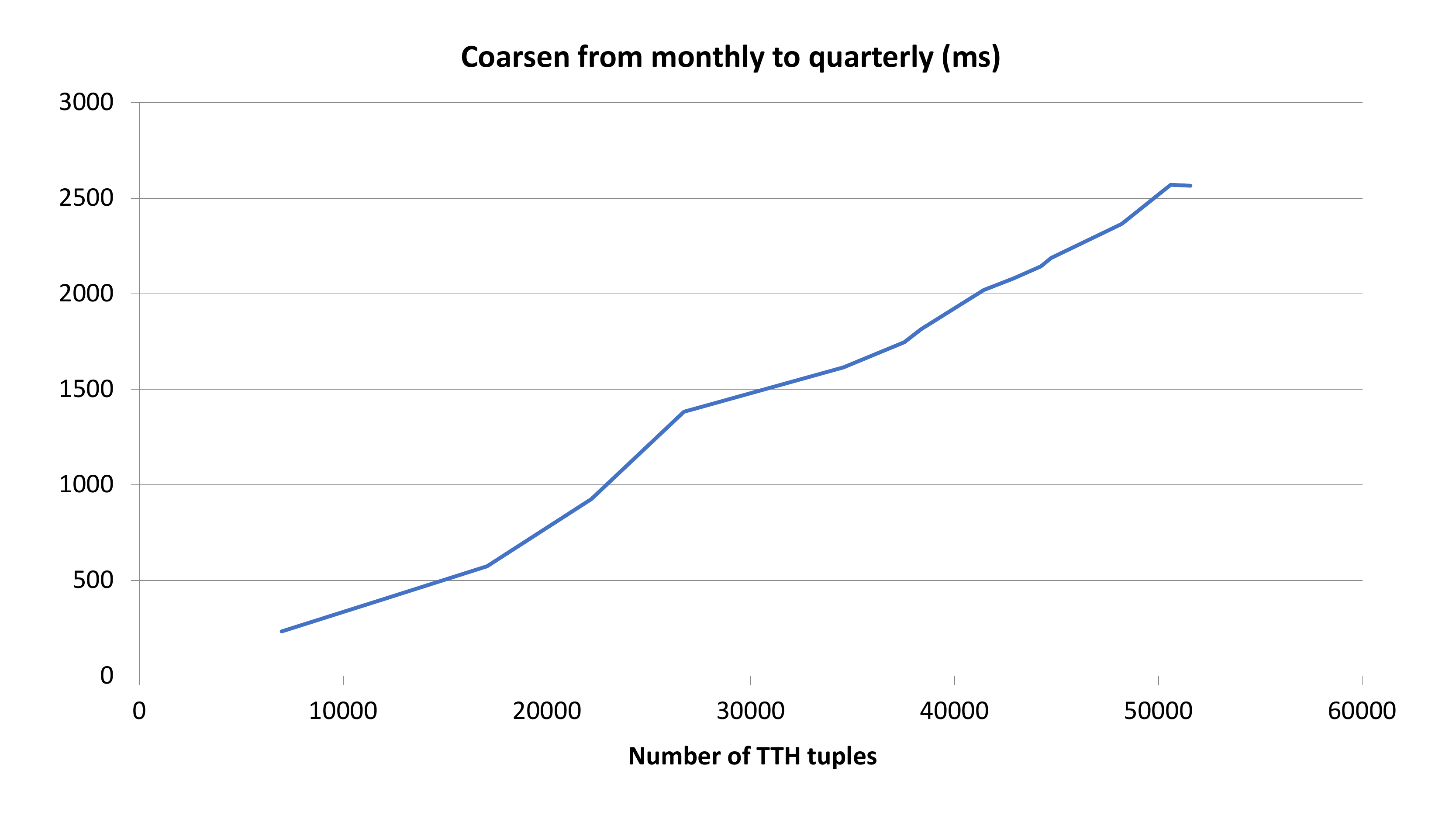}
\caption{(Left) The \textit{merge} operation over two roughly equal sized TTHs. (Right) The cost of the \textit{coarsen} operation. }
\label{fig:merge}
\end{figure*}

\noindent \textbf{Cost of the \textit{merge} Operation.} The \textit{merge} operation is effectively a bag union operation over two TTHs where the expensive part is the concatenation of the two document lists corresponding to a common term of the TTHs. In this experiment we merge the TTHs of two different newspapers by progressively inserting terms in them in batches and then merging. The number on the X axis of Figure \ref{fig:merge} (left) is the size of the merged histogram. The operation is linear and within interactive response limits.

\noindent \textbf{Cost of the \textit{coarsen} Operation.} The coarsen operation is a temporal condensation operation over TTHI tuples that reduces the number of tuples. Figure \ref{fig:merge} (right) shows that the operation is linear with respect to the size of the un-coarsened data. 

These experiments demonstrate that temporal term histograms are very useful for analytical queries are not expensive to compute, and are therefore a feasible primitive for temporal text data.
\section{Conclusion}
\label{sec:conclusion}
In this paper, we presented the temporal term histogram as an intermediate level data object that can be used for text data analytics. While we showed a how a number of analytical queries can be expressed and evaluated with TTH structures, there are indeed limitations. For example, the materialized view shown in Figure \ref{fig:histogram} can be automatically constructed only when the construction does not have arbitrary functions. For example, a query to create a materialized temporal term histogram view over all documents where the number of mentions of term \texttt{t1} is at least 10\% higher than that of \texttt{t2} cannot be generated automatically. However, if the materialized view is manually constructed for such cases, the rest of the ad hoc histogram construction and query operations can work without change. Our future work includes a more efficient implementation of the TTH by implementing the TTH functionality through the User-defined aggregate (UDA) facilities of a DBMS, and separately, by using compression techniques to reduce the size of the histogram. 


\section*{Acknowledgments}

This work has been partly funded by grant number 1738411 from the National Science Foundation.

\bibliographystyle{ACM-Reference-Format}
\bibliography{sample-bibliography}


\begin{thebibliography}{8}


\ifx \showCODEN    \undefined \def \showCODEN     #1{\unskip}     \fi
\ifx \showDOI      \undefined \def \showDOI       #1{#1}\fi
\ifx \showISBNx    \undefined \def \showISBNx     #1{\unskip}     \fi
\ifx \showISBNxiii \undefined \def \showISBNxiii  #1{\unskip}     \fi
\ifx \showISSN     \undefined \def \showISSN      #1{\unskip}     \fi
\ifx \showLCCN     \undefined \def \showLCCN      #1{\unskip}     \fi
\ifx \shownote     \undefined \def \shownote      #1{#1}          \fi
\ifx \showarticletitle \undefined \def \showarticletitle #1{#1}   \fi
\ifx \showURL      \undefined \def \showURL       {\relax}        \fi
\providecommand\bibfield[2]{#2}
\providecommand\bibinfo[2]{#2}
\providecommand\natexlab[1]{#1}
\providecommand\showeprint[2][]{arXiv:#2}

\bibitem[\protect\citeauthoryear{Gentzkow, Kelly, and Taddy}{Gentzkow
  et~al\mbox{.}}{2017}]%
        {gentzkow2017text}
\bibfield{author}{\bibinfo{person}{Matthew Gentzkow}, \bibinfo{person}{Bryan~T
  Kelly}, {and} \bibinfo{person}{Matt Taddy}.} \bibinfo{year}{2017}\natexlab{}.
\newblock \bibinfo{booktitle}{\emph{Text as data}}.
\newblock \bibinfo{type}{{T}echnical {R}eport}. \bibinfo{institution}{National
  Bureau of Economic Research}.
\newblock


\bibitem[\protect\citeauthoryear{Grimmer and Stewart}{Grimmer and
  Stewart}{2013}]%
        {grimmer2013text}
\bibfield{author}{\bibinfo{person}{Justin Grimmer} {and}
  \bibinfo{person}{Brandon~M Stewart}.} \bibinfo{year}{2013}\natexlab{}.
\newblock \showarticletitle{Text as data: The promise and pitfalls of automatic
  content analysis methods for political texts}.
\newblock \bibinfo{journal}{\emph{Political analysis}} \bibinfo{volume}{21},
  \bibinfo{number}{3} (\bibinfo{year}{2013}), \bibinfo{pages}{267--297}.
\newblock


\bibitem[\protect\citeauthoryear{Heimerl, Lohmann, Lange, and Ertl}{Heimerl
  et~al\mbox{.}}{2014}]%
        {6758829}
\bibfield{author}{\bibinfo{person}{F. Heimerl}, \bibinfo{person}{S. Lohmann},
  \bibinfo{person}{S. Lange}, {and} \bibinfo{person}{T. Ertl}.}
  \bibinfo{year}{2014}\natexlab{}.
\newblock \showarticletitle{Word Cloud Explorer: Text Analytics Based on Word
  Clouds}. In \bibinfo{booktitle}{\emph{2014 47th Hawaii International
  Conference on System Sciences}}. \bibinfo{pages}{1833--1842}.
\newblock
\showISSN{1530-1605}
\urldef\tempurl%
\url{https://doi.org/10.1109/HICSS.2014.231}
\showDOI{\tempurl}


\bibitem[\protect\citeauthoryear{Hollander, Wolfe, and Chicken}{Hollander
  et~al\mbox{.}}{2013}]%
        {hollander2013nonparametric}
\bibfield{author}{\bibinfo{person}{Myles Hollander}, \bibinfo{person}{Douglas~A
  Wolfe}, {and} \bibinfo{person}{Eric Chicken}.}
  \bibinfo{year}{2013}\natexlab{}.
\newblock \bibinfo{booktitle}{\emph{Nonparametric statistical methods}}.
  Vol.~\bibinfo{volume}{751}.
\newblock \bibinfo{publisher}{John Wiley \& Sons}.
\newblock


\bibitem[\protect\citeauthoryear{Lytras, Raghavan, and Damiani}{Lytras
  et~al\mbox{.}}{2017}]%
        {LytrasRD17}
\bibfield{author}{\bibinfo{person}{Miltiadis~D. Lytras}, \bibinfo{person}{Vijay
  Raghavan}, {and} \bibinfo{person}{Ernesto Damiani}.}
  \bibinfo{year}{2017}\natexlab{}.
\newblock \showarticletitle{Big Data and Data Analytics Research: From
  Metaphors to Value Space for Collective Wisdom in Human Decision Making and
  Smart Machines}.
\newblock \bibinfo{journal}{\emph{Int. J. Semantic Web Inf. Syst.}}
  \bibinfo{volume}{13}, \bibinfo{number}{1} (\bibinfo{year}{2017}),
  \bibinfo{pages}{1--10}.
\newblock
\urldef\tempurl%
\url{https://doi.org/10.4018/IJSWIS.2017010101}
\showDOI{\tempurl}


\bibitem[\protect\citeauthoryear{Pirzadeh, Carey, and Westmann}{Pirzadeh
  et~al\mbox{.}}{2017}]%
        {8258260}
\bibfield{author}{\bibinfo{person}{P. Pirzadeh}, \bibinfo{person}{M. Carey},
  {and} \bibinfo{person}{T. Westmann}.} \bibinfo{year}{2017}\natexlab{}.
\newblock \showarticletitle{A performance study of big data analytics
  platforms}. In \bibinfo{booktitle}{\emph{2017 IEEE International Conference
  on Big Data (Big Data)}}. \bibinfo{pages}{2911--2920}.
\newblock
\urldef\tempurl%
\url{https://doi.org/10.1109/BigData.2017.8258260}
\showDOI{\tempurl}


\bibitem[\protect\citeauthoryear{Roberts}{Roberts}{2018}]%
        {roberts2018censored}
\bibfield{author}{\bibinfo{person}{Margaret~E Roberts}.}
  \bibinfo{year}{2018}\natexlab{}.
\newblock \bibinfo{booktitle}{\emph{Censored: Distraction and Diversion Inside
  Chinas Great Firewall}}.
\newblock \bibinfo{publisher}{Princeton University Press}.
\newblock


\bibitem[\protect\citeauthoryear{Siddiqui, Lee, Kim, Xue, Yu, Zou, Guo, Liu,
  Wang, Karahalios, and Parameswaran}{Siddiqui et~al\mbox{.}}{2017}]%
        {DBLP:conf/cidr/SiddiquiLKXYZGL17}
\bibfield{author}{\bibinfo{person}{Tarique Siddiqui}, \bibinfo{person}{John
  Lee}, \bibinfo{person}{Albert Kim}, \bibinfo{person}{Edward Xue},
  \bibinfo{person}{Xiaofo Yu}, \bibinfo{person}{Sean Zou},
  \bibinfo{person}{Lijin Guo}, \bibinfo{person}{Changfeng Liu},
  \bibinfo{person}{Chaoran Wang}, \bibinfo{person}{Karrie Karahalios}, {and}
  \bibinfo{person}{Aditya~G. Parameswaran}.} \bibinfo{year}{2017}\natexlab{}.
\newblock \showarticletitle{Fast-Forwarding to Desired Visualizations with
  Zenvisage}. In \bibinfo{booktitle}{\emph{{CIDR} 2017, 8th Biennial Conference
  on Innovative Data Systems Research, Chaminade, CA, USA, January 8-11, 2017,
  Online Proceedings}}.
\newblock
\urldef\tempurl%
\url{http://cidrdb.org/cidr2017/papers/p43-siddiqui-cidr17.pdf}
\showURL{%
\tempurl}


\end{thebibliography}

\end{document}